\documentstyle[12pt,epsf]{article}

\newcommand{\be}{\begin{equation}}
\newcommand{\ee}{\end{equation}}
\newcommand{\eel}[1]{\label{#1}\end{equation}}
\newcommand{\bea}{\begin{eqnarray}}
\newcommand{\eea}{\end{eqnarray}}
\newcommand{\eeal}[1]{\label{#1}\end{eqnarray}}
\newcommand{\baq}{\begin{equation}\begin{array}{rcl}}
\newcommand{\eaq}{\end{array}\end{equation}}
\newcommand{\eaql}[1]{\end{array}\label{#1}\end{equation}}
\newcommand{\beac}{\begin{equation}\begin{array}{rcl}}
\newcommand{\eeacn}[1]{\end{array}\label{#1}\end{equation}}
\newcommand{\ba}{\begin{array}}
\newcommand{\ea}{\end{array}}
\newcommand{\non}{\nonumber \\}
\newcommand{\equ}[1]{(\ref{#1})}

\newcommand{\la}{\lambda}

\newcommand{\al}{{\alpha^{'}}}
\newcommand{\beq}{\begin{eqnarray}}
\newcommand{\eeq}{\end{eqnarray}}

\newcommand{\journal}[4]{{\rm #1~}{#2}\,(19#3)\,#4}

\newcommand{\np}{\journal {Nucl. Phys.}}
\newcommand{\pl}{\journal {Phys. Lett.}}

\begin{document}
\baselineskip 18pt
\thispagestyle{empty}

\begin{titlepage}

\begin{flushright}
{\sf TAUP--2449--97}\\
{\sf LMU-TPW-97-22}\\
{\sf hep--th/9709010}\\
{\sf September 1997}
\end{flushright}
\vfill
\begin{center}
{\bf\large On the M-Theory Approach to\\ (Compactified) 5D Field
Theories
\footnote{Research supported in part by:
the German-Israeli Foundation for Scientific Research, the Israel 
Academy of Sciences and Humanities - Centers of Excellence Program, 
the European Commission TMR Programme ERBFMRX-CT96-0045 in which 
S.T. is associated to HU-Berlin, the US-Israel
Binational Science Foundation and by the Israel Science Foundation.}}\\ 
\vskip 1cm
{\sc A. Brandhuber$^a$, N. Itzhaki$^a$,}\\
{\sc J. Sonnenschein$^a$, S. Theisen$^b$ and S. Yankielowizc$^a$}\\[5mm]

{\em $^a$School of Physics and Astronomy}\\
{\em Beverly and Raymond-Sackler Faculty of Exact Sciences}\\
{\em Tel-Aviv University, Ramat-Aviv, Tel-Aviv 69978, Israel}\\[3mm]

{\em $^b$Sektion Physik, Universit\"at M\"unchen, FRG} 

\end{center}
\vfill

\begin{abstract}

We construct M-theory curves associated with brane configurations of
$SU(N)$, $SO(N)$ and $Sp(2N)$ 5d supersymmetric gauge theories
compactified on a circle. From the curves we can account for all the
existing different $SU(N)$ field theories with $N_f \leq 2 N$.
This is the correct bound for $N \geq 3$. We remark on the 
exceptional case $SU(2)$. The  bounds obtained for 
$SO(N)$ and $Sp(2N)$ are $N_f\leq N-4$ and $N_f\leq 2N+4$,
respectively.

\end{abstract}

\end{titlepage}

\setcounter{page}{1}

{\sl Introduction}

Recently many interesting result in field theory and string theories
were obtained using branes in superstring theories. In particular,
brane configurations based on the construction of Hanany and Witten
\cite{han} led to realizations of various aspects of SUSY gauge
theories. In this note we follow \cite{ah} (see also \cite{bk})
and focus on the
brane configurations which are relevant to five dimensional N=1
supersymmetric gauge theories and five dimensional theories
compactified on a circle.

We briefly review the brane construction of 5d gauge 
theories and then 
discuss 5d theories compactified on a circle. As in
the 4d case \cite{witten} it is possible to describe these theories
using a smooth brane configuration in M theory given by a holomorphic 
curve \cite{kol}. The curves for $SU(N)$ theories can be
determined using symmetries, periodicity and requiring their
4d limits $R_B \to 0$ to be the Seiberg-Witten curves.
For $N>2$ we find agreement with field theory
results \cite{ims}. In particular, we find the same bound on $N_f$
for consistent theories and are able to identify for fixed $N$ and
$N_f$ the different theories which are classified in the field theory 
picture by the coefficient $c_{cl}$ of the Chern-Simons term in the
prepotential \cite{ims}. 
Also in the M-theory approach one does not find the field 
theories with $N_f=5,6,7,8$ for $N=2$.
The curves we find can (for some values of $c_{cl}$) be identified with 
the spectral curves of integral models which were conjectured in
ref.~\cite{nek} to be relevant for the 
non-perturbative solution of 5d theories compactified on a circle.
We also discuss the generalizations to $SO(N)$ and $Sp(2N)$ 
gauge groups.
\vskip2mm

{\sl 5D Field Theories and Type IIB description}

We begin this section with a brief review of  some of the
results for five dimensional gauge theories with eight supercharges
and simple gauge group. This analysis was initiated by Seiberg
\cite{sei2} and
generalized in \cite{msei,ims,dani}. The two possible multiplets are:  
the vector
multiplet with a real scalar in the adjoint
representation of the gauge group $G$, denoted by $\phi^a$,
and a set of hypermultiplets.
We will only discuss matter in the fundamental representation.
This might be a limitation of the brane set-up, but not of the
field theory.
The Coloumb branch is parameterized by the
scalars $\phi^i$ in the Cartan subalgebra of $G$,
$i = 1 \ldots r = {\rm rank}(G)$. The moduli space is
${\bf R}^r/{\cal W}(G)$, where ${\cal W}(G)$ is the Weyl group of $G$. 
An important quantity is the quantum prepotential
which is of the general form
\be
{\cal F} = \frac{1}{2 g_0^2} \phi^i \phi^i + \frac{c_{cl}}{6} d_{ijk} 
\phi^i \phi^j \phi^k + \frac{1}{12} \left( \sum_\alpha |\alpha_i
\phi^i|^3 - \sum_{l=1}^{N_f} \sum_{w} |w_i \phi^i + m_l|^3 \right)
\ee
The sums are over the roots and fundamental weights, respectively.
The necessary condition for the existence of a non-trivial UV fixed
point (in the limit $g_0 \to \infty$) is that the prepotential
be a convex function over the Coulomb branch. Note that the
third rank symmetric tensor $d_{ijk}$ only exist for $SU(N)$ with
$N > 2$. In all other cases $c_{cl} = 0$. We now summarize the relevant 
results of \cite{ims}: For $SU(N)$ there is a quantization
condition $c_{cl}+N_f/2\in{\bf Z}$ and only
$N_f+2|c_{cl}|\leq 2N$ are allowed. By integrating out massive
matter an effective  $c_{eff} = c_{cl} - (n_+ - n_-)/2$ is generated
where $n_+$($n_-$) denotes the number of hypermultiplets whose mass
$m_l$ was sent to $+\infty$($-\infty$).
For $SO(N)$ ($Sp(2N)$) we have $c_{cl} = 0$ and the condition is
$N_f\leq N-4$ ($N_f\leq 2N+4$).

In \cite{nek} supersymmetric 5d gauge theories compactified on a
circle of radius $R$ were studied.
The contributions to the perturbative prepotential from Kaluza-Klein states
was found, which exhibits the correct limits at small and large 
radius of the fifth dimension. 
The non-perturbative corrections were conjectured to
be encoded in the spectral curves of relativistic Toda systems. 

The brane description of $N=1$ 5D gauge
theories \cite{ah}  is related to the configurations in
\cite{han} by T-duality.
The world-volume of the NS 5-branes spans the $x^0, x^1, x^2,
x^3, x^4$ and $x^5$ directions and the D5 branes are along the
$x^0, x^1, x^2, x^3, x^4$ and $x^6$ directions.
The coordinates $x^0, x^1, x^2, x^3, x^4$ which are common
to the NS 5-branes and D5 branes are the coordinates of
the 5D field theory.
Actually, as pointed out in ref.~\cite{han}, the naive configurations 
obtained by T-duality should be reconsidered taking charge
conservation into account.
In addition to NS and D five-branes, there are, more generally, 
$(n,m)$ five-branes with tension
\be T_5^{(n, m)}=\frac{1}{(2\pi)^5\al^3}
\sqrt{\frac{n^2}{\lambda^4}+\frac{m^2}{\lambda^2}},
\ee
$n$ is the NS charge and 
$m$ the R charge; 
$\lambda$ is the IIB string coupling constant.\footnote{
This expression is valid if the RR scalar vanishes. 
This corresponds to taking a torus with rectangular unit cell 
in the next section.} 
Charge conservation implies that when a $(0,m)$ brane ends on a $(n,0)$
brane, a $(n,m)$ brane is formed. The zero force condition implies that it
satisfies \cite{ah,kol}
$\left|{\Delta x^6\over\Delta x^5}\right|=\left|{m\lambda\over n}\right|$.
This $(n,m)$ brane with this orientation in the $(x^5,x^6)$ plane
does not break the supersymmetry any further.

The brane configuration is expected to be related to the 5D gauge
theories when $\la\ll 1$. Then $(n,m)$ branes with $n\neq 0$
are much heavier then the D5 fivebranes, hence they can be treated as
a classical background.
\vskip2mm

{\sl M theory description}

The M-theory origin of the type IIB 5d theories was discussed in
\cite{kol}. In the usual type IIA picture we  start with NS branes
(012345) corresponding to unwrapped M-theory fivebranes (M5) and
D4-branes (01236) corresponding to M5-branes wrapped on $x^{10}$.
To obtain the description of the type IIB 5d theories we further
compactify the $x^4$ direction on a circle of radius $R_A$.  
Thus we consider M-theory on a torus. 
Type IIB string theory is obtained in the limit of a zero area torus
$R_A\, R_{10} \to 0$  while keeping the string coupling
$\lambda\sim\frac{R_{10}}{R_{A}}$ fixed. 
In view of the relation
\cite{schwarz} ($l_{11}$ is the 11-dimensional Planck length)
$R_B\sim\frac{l_{11}^3}{R_{10} R_A}$,
$R_B \to \infty$ in this limit; 
it becomes the new tenth type IIB dimension.
Note that in effect we are performing here the usual T-duality
transformation with $R_B\sim\alpha'/R_A$ and $\alpha'\sim l_{11}^3/R_{10}$.

In the M-theory description there is a $SL(2,{\bf Z})$ acting
on the complex structure of the two-torus on which we compactify
and thus also on the doublet of winding numbers,
$(p,q)$, of the M5 brane which wraps around the torus.
This involves the coordinates $x^4$ and $x^{10}$.
The fact that $(p,q)$ specifies the winding of the $M5$ brane
on $T^2$ as well as the orientation of the resulting
$(p,q)$ branes in the $(x^5,x^6)$ plane, means that $SL(2,{\bf Z})$
acts
on the complex coordinates $v=(x^4+i x^5)/R_A$ and
$s=(x^6+i x^{10})/R_{10}$, which transform as a doublet.

We are, therefore, considering the type IIA theory for which an M-theory
description exists with compactified $x^4$ and recover the 5d theories 
in the limit $R_A \to 0$ or, equivalently,  $R_B \to \infty$.
\vskip2mm

{\sl $SU(N)$ gauge theories}

In \cite{witten} a description of  brane configurations which are
relevant to $N=2$ gauge theories
in four dimensional space-time was given via a smooth curve in M theory
\be\label{10}F(v, t)=0,\ee
where
$t=\exp(-s/R_{10})$.
The brane configurations which we consider in the present paper are
related to the brane configuration of \cite{han} by T duality along
the $x^3,x^4$
directions, taking into account charge conservation as mentioned
in the previous section. It is possible therefore to describe them via 
a curve in M theory \cite{kol}.
Before we continue we should note that the curve which was used to
describe the four dimensional theories is smooth in the M theory limit.
To be more precise the maximal curvature (at the ends of the D4-branes 
on the NS fivebrane) is of the order of $1/R_{10}^2$ \cite{biksy}.
The M theory description of the type IIB theory is singular.
Therefore one might expect that for the brane configuration which is
relevant to describe five dimensional gauge theories the curve is
singular in the type IIB limit $R_B \to \infty$.

Indeed, using  the curve (which is defined below) one can find  that
 the maximal curvature (at the ends of the D5 branes on the NS
fivebrane) is proportional to $R_A^2R_{10}^2/(R_A^2+R_{10}^2)^3$ 
, so by going to the type
IIB limit a curvature singularity appears.
Nevertheless the M theory description will
provide some insight into the relation between the brane configurations
and 5d field theories compactified on a circle of radius $R_B$. In the 
four-dimensional case by going from the type IIA brane configuration
to the smooth brane configuration in M theory we obtained all
non-perturbative corrections.
The holomorphic curve turned out to be
the Seiberg-Witten curve and the condition that the BPS states
come from supersymmetry preserving M2 branes which end on the holomorphic
curve lead to the Seiberg-Witten differential \cite{klmvw}
For recent discussions within the context of M theory, see also
\cite{lll,FS,henning}.
Together they
contain the information about the full IR dynamics of
the N=2 gauge theory including all instanton
corrections \cite{sw}. The situation we encountered in the present
work is quite similar. If we consider the 5d case the M theory
description becomes singular and the type IIB brane configuration
receives no corrections. There are no instanton corrections 
in five dimensions \cite{sei2}.
However, as we compactify
this five dimensional theory on a circle the situation changes.
To the usual 4d one-loop prepotential, we have to add corrections
coming from an infinite tower of Kaluza-Klein states and instanton
corrections \cite{nek}. Both types of corrections are again
found by going to the M theory description where they are encoded
in the explicit form of a complex curve which is uniquely
fixed by the asymptotic behavior of the branes and the condition
of supersymmetry, i.e. holomorphic embedding in the $(v,s)$ space.
Some of the curves we find agree with the spectral curves of
relativistic Toda systems, which were
used in ref.~\cite{nek} to study the non-perturbative
solution of 5d theories compactified on a circle.
The meromorphic differential, which one also needs to  
specify the theory, is the same the on in \cite{klmvw}.

After compactifying $x^4$ on a circle, $v$ is no
longer single-valued. Following \cite{kol} we introduce
\be
w=\exp(-i\frac{v}{R_A})
\ee
and describe the curve by
\be
\label{11} F(w, t)=0.
\ee
We are interested in finding all inequivalent theories. The $S$
generator of $SL(2,{\bf Z})$ interchanges the NS5 and D5 branes. It,
thus, rotates the corresponding brane configuration by 90 degrees
taking $\lambda \to -1/\lambda$. Therefore, in the new configuration
the (new) NS5 branes will be the lighter ones and we shall consider
the effective field theory on them. This is precisely the same theory 
as the original one. We can, therefore, limit ourselves to
$SL(2,{\bf Z})$ transformations which keep the D5 branes lighter than 
the NS branes. Since we look at theories with only one gauge factor we 
restrict ourselves to configurations with two NS5 branes i.e. the
polynomial should be quadratic in $t$. This leaves
$SL(2,{\bf Z})$ transformations of the form $T^l$.
Matter is introduced via semi-infinite D5 branes.
\be
P_1(w)t^2+P_2(w)t+P_3(w)=0 ~.
\ee
Moreover, we always have the freedom to move the semi-infinite D5
branes to one side keeping the coefficient of $t^2$ normalized to one.
This can be achieved by the transformation
$t\rightarrow P_3(w)/t$ (such a transformation moves all
semi-infinite four branes to the right hand side of the NS branes) which 
yields
\be
t^2 + P_2(w) t + P_4(w) = 0,
\eel{su-curve}
where $P_4=P_1 P_3$.
If we express the curves as vanishing conditions in $w$ and $t$,
as opposed to $v$ and $t$, we have to take into account that
$w=\infty$ and $w=0$ correspond to the asymptotic region
(whereas $v=0$ does not). That means that the multiplicity of the
zero roots of the polynomials $P_2$ and $P_4$ will be relevant.
We therefore write them in the following general form
\be 
P_2(w)=c_2\, w^n\prod_{i=1}^{N}(w-\tilde a_i)\,,
\;\;\;P_4(w)=c_4\, w^m\prod_{j=1}^{N_f}(w-\tilde m_j).
\eel{gencurve}
The integers $n$, $m$ characterize the
underlying field theory, as will become explicit below.
To specify the field theory which is described by the brane
configuration associated with a given curve we need to study the
asymptotic behavior of the branes.

Let us first consider $w \rightarrow\infty$. To leading order we get
(after a rescaling)
\be
t^2+c w^{N+n}t+w^{N_f+m}=0.
\ee
The asymptotic behavior is therefore
\beq
 t_1\sim w^{N+n},\;\; t_2\sim w^{N_f+m-N-n} && \mbox{when}\;\;\;\; 
2(N+n)>N_f+m,\non
t_{1, 2}\sim w^{N+n} && \mbox{when}\;\;\;\;
2(N+n)=N_f+m,\\
t_{1, 2}\sim w^{(N_f+m)/2} && \mbox{when}\;\;\;\;
2(N+n)<N_f+m.\nonumber
\eeq
For $2(N+n)<N_f+m$ the asymptotic behavior of the branes depends
only on $N_f$. The type IIB description of such theories leads to
crossing of the NS fivebranes and hence new massless excitations will
appear.
The analog case in \cite{witten} leads to gauge theories with
positive  beta function.
In the case of finite $R_B$ no singularity associated with brane  
crossing appears in the curve. The branes do not really cross but they  
approach each other asymptotically. It would be interesting to  
further investigate this range since it may lead to new theories as  
was conjectured by \cite{ah}. One does not expect these theories to be  
$SU(N)$ field theories with $N_f$ flavors since those do not lead to  
non-trivial fixed points  for $N_f > 2 N$ ($N \geq 3$) \cite{ims}.
For ordinary $SU(N)$ theories we should focus on
$2(N+n)\geq N_f+m$. In four dimensions they correspond to
asymptotically free theories, whereas in  five dimensions they will
lead to non-trivial IR fixed points. The case $2(N+n) = N_f+m$
corresponds to a situation where in the IIB picture there are two
parallel five branes which go off to $x^5=+\infty$ at a finite
distance apart. In the 4d limit this corresponds to the superconformal 
theories where the distance between the parallel fivebranes amounts to 
a choice of the gauge coupling $\tau$.

Consider now $w\rightarrow 0$ ($x^5 \to -\infty$) the leading order
polynomial is
\be
t^2+c' w^nt+w^m=0
\ee
and the asymptotic behavior is
\beq
t_{1, 2}\sim w^{m/2} &&\mbox{when} \;\;\;\; 2n>m,\non
t_{1, 2}\sim w^n &&\mbox{when} \;\;\;\; 2n=m,\\
t_1\sim w^n,\;\;t_2\sim w^{m-n} &&\mbox{when} \;\;\;\; 2n<m.\nonumber
\eeq
Curves with $2n>m$  correspond to type IIB configurations with crossing 
semi-infinite fivebranes (or asymptotically approaching fivebranes for 
finite $R_B$).
Thus we concentrate on curves with $2n\leq m$.
As mentioned above, the only $SL(2,{\bf Z})$ transformations
which are not
yet fixed by the Ansatz \equ{su-curve} are $T^l$.
Such a transformation acts as
$t\rightarrow tw^{l}, \;\;w\rightarrow w$ which takes
$m\rightarrow m+2l$ and $n\rightarrow n+l$.
Since $2n\leq m$ one can set $n=0,\,m\geq 0$.
We are thus left with the single constraint
\be
2N\geq N_f+m.
\ee
Which means that $m=0, 1,\dots,2N-N_f$.
The parity operator $w\rightarrow 1/w$ ($v\rightarrow -v$) takes
\be
\label{what} m\rightarrow 2N-N_f-m.
\ee
Therefore the number of allowed values for $m$ which yields different 
curves is
$[(2N-N_f+1)/2]$.
For $N>2$ this result is in agreement with the field theory result
\cite{ims}
where it was found that the number of allowed values for $|c_{cl}|$
is $[(2N-N_f+1)/2]$. Recall that $c_{cl}$ is the numerical
coefficient of the bare cubic term in the prepotential which
characterizes the theory.
We have thus found that the brane configurations we have obtained from 
M theory simply reproduce the known superconformal field theories.

To find the relation between $|c_{cl}|$ and $m$ we note that the parity 
transformation $w\rightarrow 1/w$ acts as charge conjugation as it 
reverses the orientation of the elementary strings.
Since charge conjugation takes $c_{cl}\rightarrow -c_{cl}$ \cite{ims} 
we identify, using eq.(\ref{what})
\be
c_{cl} = N - m - N_f/2.
\ee
Below we show, using the M theory approach,
that the brane description agrees with the field theory result
\cite{ims} (this was already shown using the type IIB description
\cite{bk}).
The one exceptional case is  $SU(2)$ were it was argued that theories 
with $N_f=5, 6, 7$ are also consistent \cite{sei2} while no consistent 
brane configuration could be constructed \cite{ah}.

The $SU(2)$ theories with $N_f = 5,6,7$ correspond to interacting fixed 
points with global symmetry groups $E_6,E_7,E_8$ \cite{sei2}. $N_f=8$
is also consistent but does not lead to an interesting fixed point.
But these theories are outside the range $N_f \leq 2 N$ of allowed
brane configuration. In ref.~\cite{ah} it was explained that by going 
beyond this bound two semi-infinite branes necessarily cross and this 
crossing would induce additional massless states. In our M theory
configurations \equ{su-curve} this singularity does not occur but the 
two branes approach each other asymptotically. Nevertheless also in
this setup we expect additional light  states which have no
conventional field theory interpretation.

After having given the general form for the curves in eq.\equ{gencurve}, 
we will now rewrite them in a form that allows us to go to the $d=4$ limit by
taking $R_B\to 0$ or, equivalently, $R_A\to\infty$. 
By an appropriate choice of the constants $c_2$ and $c_4$ one obtains
\be
t^2+2t w^{N/2}\prod_{i=1}^{N}R_A\sin\Bigl({v - a_i\over 2R_A}\Bigr)+
w^{N_f/2+m}\prod_{j=1}^{N_f} R_A \sin\Bigl({v - m_j\over 2R_A}\Bigr)=0.
\eel{su-curve1}
$a_i$ are the positions of the finite D branes in $v$ space and 
$m_i$, the bare masses of the hypermultiplets,  are the positions of the 
semi-infinte D branes. 
Note that whereas in the five-dimensional theories the masses are real, 
they are complex in the compactified theory. 
The parameters are related to those in 
eq.\equ{gencurve} via $\tilde a_i=e^{-a_i/R_A},\, \tilde m_i=e^{-i m_i/R_A}$.
For $R_A \to \infty$ or $R_B \to 0$ this becomes
\be
t^2 + 2 t \prod_{i=1}^{N} (v - a_i) + \prod_{j=1}^{N_f} (v - m_j) = 0 .
\ee
which agrees with the  curves for $SU(N)$ N=2 SQCD with $N_f$ flavors 
of \cite{mw,mmm}

On the other hand we can study the large $R_B$ region of these curves 
and integrate out hypermultiplets by sending their masses to infinity. 
There are two possibilities $m \to \pm \infty$ which leads to two
different curves which correspond  to two theories with equal matter
content but different values of $c_{cl}$. Starting from the unique
curve for $N_f = 2 N$ the flow pattern of the curves reproduces all 
possible theories labeled by $c_{cl}$. Also the relation of $c_{cl}$ 
to the number of quarks with positive and negative masses has an
explanation in the M theory picture.

In the 4d case one can obtain the effective gauge coupling by
considering the asymptotic bending of the NS branes which can be read 
directly from the associated M-theory curve. This does not hold in
the 5d case. It is reflected by the fact that calculating
$\log(t_1/t_2)$ where $t_{1,2}$ are the two roots associated with the 
curve \equ{su-curve1} does not give the effective gauge coupling in
the limit $R_B \to \infty$. This bending calculation reflects, however, 
the $N_f \leq 2 N$ bound. The correct way to define and obtain the 5d 
effective gauge coupling is by calculating some BPS mass as is
explicitly demonstrated in ref.~\cite{ah}. It would be important to
rederive it geometrically via the M-theory approach.
\vskip2mm

{\sl $SO(N)$ and $Sp(2N)$ gauge theories}

The M theory description of $SO(N)$ and $Sp(2N)$ gauge theories in four
dimensions was given in \cite{lll,bsty2}.
A subtlety in $d=4$ was how the orientifold planes, which are present in the 
type IIA formulation would appear in the M theory description. 
Here we will determine curves which have the correct behavior
in the four dimensional limit and which respect the symmetries, 
i.e. they must be symmetric under \footnote{As in the 
four-dimensional case, for $SO(2n+1)$ we must at the same time 
transform $t\to -t$; cf.\cite{bsty2}. Also, a shift $v\to v+2\pi R_A$
must be accompanied by a shift $s\to s+\pi R_{10}$.} $w\to 1/w$,          
corresponding to $v\to -v$, 
and be periodic in $v$. We first note that the first of these 
conditions does not allow for the introduction of the parameter $m$ which 
distinguished different theories in the $SU(N)$ case. 
For $SO(2N)$ we thus arrive at the curve
\bea
t^2\left(\sin\frac{v}{R_A}\right)^2 \!\!\!\! & + & \!\!\!\!2 t \prod_{i=1}^{N}
\sin\left(\frac{v-a_i}{2R_A}\right)\sin\left(\frac{v+a_i}{2R_A}
\right)\non
+ \!\!\!\!& c &\!\!\!\! \left(\sin\frac{v}{R_A}\right)^2\prod_{j=1}^{N_f}
\sin\left(\frac{v-m_j}{2R_A}\right)\sin\left(\frac{v+m_j}{R_A}
\right) = 0
\eea
By appropriate choice of $c$ and rescaling of $t$ one can take
the limit $R_A\to\infty$ and arrives at the well known 
Seiberg-Witten curves for $SO(2N)$ with $N_f$ fundamental 
hypermultiplets. 

Note that $m_i$ and $-m_i$ enter in the curve symmetrically due to the 
reflection symmetry $v \to -v$. In particular, this implies that
$c_{cl}=0$ in this case since the flows $m \to +\infty$ and
$m \to -\infty$ are equivalent (in contrast to the $SU(N)$ case).
This is also in agreement with field theory results \cite{ims}.
This will also carry over to the $SO(2N+1)$ and $Sp(2N)$ 
theories to be discussed below. 
This curve has the expected behavior as we go to $R_A \to \infty$
i.e. it turns into the four-dimensional curve (after
a rescaling of $t$ and $v$). On the other hand if we investigate the
curve for large $R_B$ and study the behavior for $|v| >> |a_i|, |m_j|$ 
one should reproduce the bending (tilting) of the five dimensional
brane configurations. Indeed we find
\be
t^2\exp(2|v|/R_A)+t\exp(N|v|/R_A)+\exp((2+N_f)|v|/R_A)=0
\ee
For 
\be
2N-4\ge N_f ~.
\eel{socond}
one finds, 
\be
\log(t_1/t_2) \sim R_B (4 N - 8 - 2 N_f) |v| ~,
\ee
i.e. in the asymptotic region there are two  
branes which diverge from each other or move off to 
infinity at a finite distance. For $4 N - 8 < 2 N_f$, on the other 
hand, they asymptotically approach one another.
The condition \equ{socond} agrees with ref.~\cite{bk}. 
Note that it differs
from the condition for asymptotically free $SO(2N)$ theories in four 
dimensions ($2N-2>N_f$).

The discussion for $SO(2N+1)$ and $Sp(2N)$ is analogous, so we will be 
very brief. 

For $Sp(2N)$ gauge groups the curve is:
\bea
t^2 + 2 t \left(\sin\frac{v}{R_A}\right)^2
\prod_{i=1}^{N}\!\!\!\!\!\!& &\!\!\!\!\!\sin\left(\frac{v-a_i}{2R_A}\right)
\sin\left(\frac{v + a_i}{2R_A}\right) \non
& + & c\prod_{j=1}^{N_f} \sin\left(\frac{v - m_j}{2R_A}\right)
\sin\left(\frac{v + m_j}{2R_A}\right) = 0
\eea
The allowed range of theories is now $2N+4\ge N_f$,
in agreement with field theory results \cite{ims}.

Finally, the $SO(2N+1)$ curve follows from the $SO(2N)$ curve by 
realizing that one of the $a_i$ should vanish and that this zero 
should be simple. This gives  
\bea
t^2 \left(\sin\frac{v}{R_A}\right)^2 \!\!\!\! & + & \!\!\!\!2t 
\sin\left(\frac{v}{2R_A}\right)\prod_{i=1}^{N}
\sin\left(\frac{v-a_i}{2R_A}\right)\sin\left(\frac{v+a_i}{2R_A}
\right)  \non
+ \!\!\!\!& c &\!\!\!\! \left(\sin\frac{v}{R_A}\right)^2\,\prod_{j=1}^{N_f}
\sin\left(\frac{v-m_j}{2R_A}\right)\sin\left(\frac{v+m_j}{2R_A}
\right) = 0
\eea
which leads to the expected bound $2N-3 \ge N_f$. 
\vskip2mm

Note that the limits on $N_f$ for orthogonal and symplectic 
groups can be interpreted from the brane configuration in the same 
way as in the four dimensional case if one takes into account that
an $O5$ plane now has (minus) the charge of a $D5$ plane, so that 
by the combined arguments of refs.~\cite{ejs,egkrs,bsty2} one   
needs to add twice as many semi-infinite ($SO$) or finite ($Sp$)
D5 branes at the position of the orientifold plane.

{\sl Summary}

The M-theory approach can be extended to discuss type
IIB configurations and their corresponding (compactified) 5d field  
theories. From the M-theory point of view we still have just one
M5 brane embedded in 
${\bf R}^{3,1}\times{\bf R}^2 \times {\bf T}^2$ where the torus corresponds  
the $(x^4, x^{10})$ subspace. We have constructed the curves which  
account for $SU(N)$, $SO(N)$ and $Sp(2N)$ 5d supersymmetric gauge  
theories compactified on a circle. In particular, for the $SU(N)$  
case we account for all 5d theories with $N_f \leq 2 N$ and identify  
the parameter in the curve corresponding to $c_{cl}$. Recall that  
from the field theory point of view it is $c_{cl}$ that characterizes   
the theory \cite{ims}.
Here we find the curves associated with these theories. The bound  
$N_f \leq 2 N$ is correct for $N \geq 3$. However for  
$N = 2$ it is known \cite{sei2} that there exist theories for $N_f  
=5, 6, 7$ which lead to non-trivial fixed points with exceptional  
global symmetries $E_{6,7,8}$. We note that for these theories  
also no brane construction is known (but they can be realized using  
branes as probes \cite{sei2}). It seems that whenever there are  
theories with exceptional symmetries it is difficult to get them using  
flat brane constructions or in the related M-theory approach.  
These theories can however be 
constructed within the geometric engineering approach in which  
a compactification on some curved (non-compact) manifold is  
considered \cite{kkv}.  

The curves which we have found are intimately related to the curves  
which were considered in the discussion of integrable 5d models  
compactified on a circle \cite{nek}. In our analysis we have  
considered only curves leading to theories with classical gauge groups  
and $N_f$ flavors. We have found in all cases the known bounds on  
$N_f$. In the analysis of the different inequivalent curves we have  
discarded curves which do not lead to such theories. It corresponds in  
the brane picture to brane configurations with $N_f$ outside this  
bound ($N_f \geq 2 N$ for the $SU(N)$ case) which necessarily  
involve more intersections of the fivebranes than the ones which exist  
on the Coulomb branch within the bound. It would be important to  
further investigate these curves and see whether they correspond to new  
interesting superconformal theories in the IR limit as was  
conjectured in \cite{ah}.

We want to close with a comment on an alternative way of introducing matter
in the five dimensional systems.
For the brane description of three and four dimensional gauge theories
this was also possible via infinite 
D5 and D6 branes, respectively. This way of introducing fundamental 
matter multiplets allows for the discussion of the Higgs branches of these 
theories. One might now try to extend this to the five-dimensional 
theories discussed in this letter and arrive at the appropriate brane 
configuration via T duality. This is however not straightforward, for the 
following reasons. First we recall that the type IIB brane configurations
discussed here are not the naive T duals of the ones discussed in 
$d=3$ and $d=4$, since T duality would not automatically lead to the 
polymeric brane configurations, but rather to a network of straight
branes. Also, the five and six branes necessary to discuss the Higgs 
branches in $d=3$ and $d=4$ would, under T duality, turn into D7 branes, 
which, since their transverse space-time is 2+1 dimensional, 
would be expected to lead to a 
deficit angle. Indeed, the 7 branes constructed in \cite{GGP}
give a {\it constant} deficit angle of ${\pi\over 6}$. However, 
they are not related to lower dimensional D branes via T-duality. 
For IIB theory compactified on a circle, there exists a seven brane 
solution which is T-dual to the six brane of IIA \cite{BdRGPT}. It reduces, 
however, in the decompactification limit to flat 10 dimensional 
Minkowski space-time, i.e. there is no deficit angle. 
This issue needs further study.

\vskip3mm

We thank the Max-Planck-Institute in Munich and the Theory
Division of CERN for hospitality.


\begin{thebibliography}{299}

\newcommand{\bi}[1]{\bibitem{#1}}

\bi{han} 
A. Hanany and E. Witten,
\np{B492}{97}{152}, hep-th/9611230

\bi{ah} 
O. Aharony and A. Hanany,
``Branes, Superpotentials and Superconformal Fixed Points",
 hep-th/9704170

\bi{bk} 
I. Brunner and A. Karch,
``Branes and Six Dimensional Fixed Points",
hep-th/9705022

\bi{witten} 
E. Witten,
``Solutions of Four-Dimensional Field Theories via $M$
Theory",
hep-th/9703166

\bi{kol} 
B. Kol,
``5d Field Theories and M Theory",
hep-th/9705031

\bi{ims} 
K. Intriligator, D. R. Morrison and N. Seiberg, \\ 
\np{B497}{97}{56}, hep-th/9702198

\bi{nek} 
N. Nekrasov, ``Five Dimensional Gauge Theories and Relativistic 
Integrable Systems'', 
hep-th/9609219

\bi{sei2} 
N. Seiberg,
\pl{388B}{96}{753}, hep-th/9608111

\bi{msei} 
D. R. Morrison and N. Seiberg,
\np{B483}{97}{229}, hep-th/9609070

\bi{dani} 
U. H. Danielsson, G. Ferretti, J. Kalkkinen and P. Stjernberg,
Phys. Lett. B405 (1997) 265, hep-th/9703098

\bi{schwarz} 
J. H. Schwarz, ``Lectures on Superstring and M Theory Dualities'',
hep-th/9607201

\bi{biksy} 
A. Brandhuber, N. Itzhaki, V. Kaplunovsky, J. Sonnenschein
and S. Yankielowicz, 
``Comments on the M Theory Approach to N=1 SQCD and Brane Dynamics'',
hep-th/9706127

\bi{klmvw} 
A. Klemm, W. Lerche, P. Mayr, C. Vafa and N. Warner,
\np{B477}{96}{746}, hep-th/9604034

\bi{lll} 
K. Landsteiner, E. Lopez and D. A. Lowe,
``N=2 Supersymmetric Gauge Theories, Branes and Orientifolds",
hep-th/9705199

\bi{FS} 
A. Fayyazuddin and M. Spalinski,  
``The Seiberg-Witten Differential from M Theory'', 
hep-th/9706087

\bi{henning} 
M. Henningson and P. Yi, hep-th 9707251

\bi{sw} 
N. Seiberg and E. Witten, \np{B426}{94}{19}, hep-th/9407087;
\np{B431}{94}{484}, hep-th/9408099

\bi{mw} 
E. Martinec and N. Warner, Nucl. Phys. B459 (1996) 97, hep-th/9509161

\bi{mmm} 
A. Marshakov, A. Mironov and A. Morozov, 
``More Evidence for the WDVV Equations in N=2 SUSY Yang-Mills Theories'', 
hep-th/9701123; A. Mironov, ``MDVV Equations in Seiberg-Witten 
Theory and Associative Algebras'', hep-th/9704205

\bi{ejs} 
N. Evans, C.V. Johnson and A. D. Shapere,
``Orientifolds, Branes, and Duality of 4D Gauge Theories'',
hep-th/9703210

\bi{bsty2} 
A. Brandhuber, J. Sonnenschein, S. Theisen and S. Yankielowicz,
``M Theory and Seiberg-Witten Curves: Orthogonal and Symplectic
Groups",
hep-th/9705232

\bi{kkv} 
S. Katz, A. Klemm and C. Vafa, Nucl. Phys. B497 (1997) 173,  
hep-th/9609239; 
S. Katz, P. Mayr and C. Vafa, 
``Mirror Symmetry and Exact Solution of 4-D N=2 Gauge Theories: 1.'',
hep-th/9706110

\bi{GGP} 
G. Gibbons, M. Green and M. Perry,
Phys. Lett. B370 (1996) 37, 
hep-th/9511080 

\bi{BdRGPT} 
E. Bergshoeff, M. de Roo, M. Green, G. Papadopoulos and P. Townsend, 
Nucl. Phys. B470 (1996),
hep-th/9601150

\bi{egkrs}
S. Elitzur, A. Giveon, D. Kutasov, E. Rabinovici and
A. Schwimmer,
``Brane Dynamics and N=1 Supersymmetric Gauge Theory",
hep-th/9704104

\end{thebibliography}
\end{document}